\documentclass{article} 

\newcommand{\eprint}{\textsf} 

\newtoks\reportnoregister \newtoks\eprintnoregister
\newcommand{\reportnumber}[1]{\reportnoregister={#1}}
\newcommand{\eprintnumber}[1]{\eprintnoregister={#1}}

\reportnumber{\mbox{}} 
\eprintnumber{\mbox{}} 

\newcommand{\reportid}{
   \begin{minipage}{17cm}\vspace{-3.2cm}
     \begin{flushright}
      {\normalsize \the\reportnoregister \\[-.2cm]
            \eprint{\the\eprintnoregister}}\vspace{3.2cm}
     \end{flushright}
   \end{minipage}\hspace{-17cm} }

\catcode`@=11   
\def\title#1{\gdef\@title{\reportid#1}}
\catcode`@=12   
\newcommand{\mechgen}{03.20.+i}      
\newcommand{\symmcons}{11.30.-j}     

\def\journalfont{\rm}         
\def\jou#1{{\journalfont #1\ }}
\def\joudef#1#2{\def #1{\jou{\ignorespaces #2}}}

\joudef{\aaa}    { Astron.\ Astrophys.}
\joudef{\aip}    { Adv.\ Phys.}
\joudef{\adm}    { adv.\ math.}
\joudef{\am}     { Ann.\ Math.}
\joudef{\apny}   { Ann.\ Phys.\ (N.Y.)}
\joudef{\apj}    { Astrophys.\ J.}
\joudef{\cjp}    { Can.\ J.\ Phys.}
\joudef{\cmp}    { Commun.\ Math.\ Phys.}
\joudef{\cqg}    { Class.\ Quantum Grav.}
\joudef{\grg}    { Gen.\ Rel.\ Grav.}
\joudef{\ijmpd}  { Int.\ J.\ Mod.\ Phys.\ D}
\joudef{\ijtp}   { Int.\ J.\ Theor.\ Phys.}
\joudef{\invm}   { Invent.\ Math.}
\joudef{\jm}     { J.\ Math.}
\joudef{\jmaa}   { J.\ Math.\ Anal.\ Appl.}
\joudef{\jmp}    { J.\ Math.\ Phys.}
\joudef{\jpa}    { J.\ Phys.\ A}
\joudef{\mnras}  { Mon.\ Not.\ R.\ Ast.\ Soc.}
\joudef{\mpla}   { Mod.\ Phys.\ Lett.\ A} 
\joudef{\nature} { Nature}
\joudef{\nc}     { Nuovo Cim.}
\joudef{\npb}    { Nuc.\ Phys.\ B}
\joudef{\ph}     { Physica}
\joudef{\pla}    { Phys.\ Lett. A}
\joudef{\plb}    { Phys.\ Lett. B}
\joudef{\pr}     { Phys.\ Rev.}
\joudef{\prd}    { Phys.\ Rev.\ D}
\joudef{\prep}   { Phys.\ Rep.}
\joudef{\prl}    { Phys.\ Rev.\ Lett.}
\joudef{\prsla}  { Proc.\ Roy.\ Soc.\ Lond.\ A}
\joudef{\ptp}    { Prog.\ Theor.\ Phys.}
\joudef{\ptps}   { Prog.\ Theor.\ Phys.\ Suppl.}
\joudef\rmp      { Rev.\ Mod.\ Phys.}
\joudef\spj      { Sov.\ Phys.\ JETP}

\usepackage{amsmath,amssymb}
\newcommand{\ncmnd}{\newcommand}
\ncmnd{\nms}{\negmedspace}
\ncmnd{\nts}{\negthickspace}
\ncmnd{\mcl}[1]{\mathcal{#1}}
\ncmnd{\beq} {\begin{equation}}
\ncmnd{\eeq} {\end{equation}}
\ncmnd{\rarr} {\rightarrow}
\ncmnd{\larr} {\leftarrow}
\ncmnd{\lrarr} {\leftrightarrow}
\ncmnd{\lbeq}[1]  {\label{eq: #1}}
\ncmnd{\refeq}[1] {(\ref{eq: #1})}
\ncmnd{\mrm}    {\mathrm}
\ncmnd{\ga}{\Gamma}
\ncmnd{\si}{\Sigma}
\ncmnd{\back}{\!\!\!\!\!\!\!\!\!\!\!}
\ncmnd{\nn}{\nonumber}
\ncmnd{\mbf}[1] {{\mathbf #1}}
\ncmnd{\bbet}{\bar{\beta}}
\ncmnd\sgn{\mathop{\rm sgn}\nolimits}
\ncmnd\real{\mathop{\rm Re}\nolimits}
\ncmnd\imag{\mathop{\rm Im}\nolimits}
\ncmnd{\re}{\Re{e}} 
\ncmnd\ts\textstyle

\begin{document}

\reportnumber{SUITP 98-21}
\eprintnumber{solv-int/9811005}

\title{A unified treatment of cubic invariants at fixed and arbitrary energy}
\author{Max Karlovini\footnote{E-mail: \eprint{max@physto.se}}\; 
        and
        Kjell Rosquist\footnote{E-mail: \eprint{kr@physto.se}} \\[10pt]
        {\small Department of Physics, Stockholm University}  \\
        {\small Box 6730, 113 85 Stockholm, Sweden} }
\date{}
\maketitle

\vspace{4cm}

\begin{abstract}{\normalsize
Cubic invariants for two-dimensional Hamiltonian
systems are investigated using the Jacobi geometrization
procedure. This approach allows for a unified treatment of invariants
at both fixed and arbitrary energy. In the geometric picture the
invariant generally corresponds to a third rank Killing tensor, whose
existence at a fixed energy value forces the metric to satisfy a
nonlinear integrability condition expressed in terms of a K\"{a}hler 
potential. Further conditions, leading to a system of equations which is 
overdetermined except for singular cases, are added when the energy is 
arbitrary. As solutions to these equations we obtain several new
superintegrable cases in addition to the previously known cases. We
also discover a superintegrable case where the cubic invariant is of a
new type which can be represented by an energy dependent linear
invariant. A complete list of all known systems which admit a cubic
invariant at arbitrary energy is given.
}\end{abstract}
\vspace{2cm}
\centerline{\bigskip\noindent PACS numbers: \mechgen \quad \symmcons }
\clearpage


\section{Introduction}

The quest for integrable systems has a long history going back more
than a century.  However, to this day, no general systematic method
exists for finding the invariants of a given system.  An effective
criterion to test if a linear invariant exists in a 2-dimensional
system was found only recently in \cite{u/b/m:classein} (compare also
\cite{rosquist/goliath:mini}). In the absence of such
general methods one can at least study and classify sufficiently
simple integrable systems with certain simple types of invariants. In
the present work, we focus on cubic invariants admitted by natural
Hamiltonian systems in two-dimensional Euclidean space. 

Cubic invariants in two dimensions have been studied before by several
workers. In 1935 Drach \cite{drach:invcubic} carried out the first
systematic study of such systems. However, using null variables
without restricting them to be complex conjugate, Drach's Hamiltonians
were not constrained to be the sum of the Euclidean space kinetic
energy and a real potential. This is probably why his results have not
often been compared to those of more recent studies such as that of
Holt in 1982 \cite{holt:cubinv}. The most complete classification of
2-dimensional systems admitting a cubic invariant, but not including
the Drach systems, was later given by Hietarinta in 1987
\cite{hietarinta:secinv}.  Most workers have dealt with arbitraray
energy (strongly conserved) invariants, but fixed energy (weakly
conserved) invariants have also been investigated to some extent (c.f.
\cite{hietarinta:secinv}).  In this paper we use the Jacobi
geometrization method to represent the dynamics.  That approach was
used in \cite{rp:inv} to give a unified treatment of quadratic
invariants at fixed and arbitrary energy. We will show that such a
unification is possible also for systems with cubic second invariants.

The Jacobi formulation has the advantage of encoding the entire
dynamics in a single geometric object, the Jacobi metric.  The system
orbits are simply the geodesics of the Jacobi geometry.  To analyze
the system one can utilize all the mathematical and computer algebra
tools developed for Riemannian geometry. Other geometric methods which
have been used for integrable systems can be found, e.g., in
\cite{per:intlie}.  In particular we mention the projection method of
Olshanetsky and Perelomov \cite{op:projintro}. That formulation,
however, is fundamentally different from the Jacobi geometrization.
For example, the projection geometry itself is not sufficient to
represent the dynamics. On the other hand the projection geometry is
designed to be very simple, typically a space of constant curvature.
A drawback is that there is no general algorithm to find a suitable
projection.

A polynomial invariant of the geodesic equations corresponds to a
single geometric object on the configuration space, known as a Killing
tensor.  The vanishing of the Poisson commutator of the invariant and
the Hamiltonian gives a set of conditions which are referred to as the
Killing tensor equations. Because of Noether's theorem, such
invariants correspond to certain dynamical symmetries of the geodesic
equations \cite{rosquist:ktsym}.  When dealing with cubic invariants,
in contrast to the case of quadratic invariants, it turns out to be
very natural to use a K\"ahler potential for the Jacobi metric.  We
find that the condition imposed on the geometry by the Killing tensor
equations is a nonlinear PDE in the K\"ahler potential, whereas the
corresponding condition for the second rank case is a linear PDE
\cite{rp:inv}. Imposing the additional condition of invariance at
arbitrary energy, the nonlinear PDE splits into a system of three
equations which in general is overdetermined. This accounts for the
fact that only isolated cases of systems with a cubic invariant at
arbitrary energy are known.  Comparing this method with the more
direct approach due to Holt \cite{holt:cubinv,hietarinta:secinv}, we
find the two final sets of equations to be very different. In
particular, while Holt's approach leads to a PDE whose nonlinearity
has no a priori restriction, our nonlinear PDE is always quadratic.
Although our work therefore represents an improvement, the equations
are nevertheless complicated and in practice impossible to solve in
full generality.  The fruitfulness of our particular approach is in
any case confirmed by the fact that we are able to present a number of
new results concerning systems which are superintegrable (i.e.
admitting the maximal number of independent invariants) at arbitrary
energy.  Furthermore we obtain all previously known strongly conserved
cubic invariants, and are thus able to give a unified classification
of all known cases.

In addition, we have also considered strongly conserved cubic
invariants of a new type which correspond to Killing vectors rather
then Killing tensors of rank three. The existence of such
``quasi-linear'' invariants seems to be a new feature.  One nontrivial
superintegrable system of this type is presented, and we also show how
its potential via coupling constant metamorphosis and coordinate
translation is in fact dual to the harmonic oscillator potential.

\section{Jacobi geometry and Killing tensors}

The Jacobi geometrization procedure relies on the fact that given a
Hamiltonian of the classical type
\begin{equation}\lbeq{hamclass}
  \textstyle H = T + V, \quad T=\frac12 h^{\alpha\beta}p_\alpha
  p_\beta, \quad V=V(q), 
\end{equation}
the orbits on a fixed energy surface $H=E$ can be mapped onto
geodesics of the Jacobi metric 
\begin{equation}
  g_{\alpha\beta} := 2(E-V)h_{\alpha\beta}
\end{equation}
on the fixed energy surface
\begin{equation} 
  H_J := \ts\frac12g^{\alpha\beta}p_\alpha p_\beta = \frac12, \quad
  g^{\alpha\gamma}g_{\gamma\beta} = \delta^\alpha{}_\beta.
\end{equation}
The mapping is achieved via the time reparametrization $t\rarr t_J$
where $dt_J = 2(E-V)dt$. Thus the Hamiltonians $H$ and $H_J$
represent the same system but in two different time gauges, often
referred to as the physical and the Jacobi time gauge,
respectively. For a less sketchy discussion of this time
reparametrization, the reader is referred to \cite{rp:inv}. 

As will be made clear below a physical polynomial invariant $I$ always
corresponds to a geodesic invariant $I_J$ which is a
\emph{homogeneous} polynomial:
\begin{equation}
  I_J = K^{\mu_1\cdots\mu_m}p_{\mu_1} \cdots p_{\mu_m},
\end{equation}
where $K^{\mu_1\cdots\mu_m}$ is a symmetric tensor on the
configuration space, with the tensor rank $m$ in general being the
same as the polynomial degree of $I$. For the case when $I$ is cubic,
the correspondence between $I$ and $I_J$ will be given explicitly in
section \ref{section:mapping}, and the method used there can also be
applied to polynomial invariants of any other degree. The equation
of motion $\{I_J,H_J\}=0$ can readily be shown to be equivalent
to the covariant equation 
\begin{equation}
  K_{(\mu_1\cdots\mu_m;\mu_{m+1})}=0,
\end{equation}
where the parenthesis denotes symmetrization, the semi-colon denotes
covariant derivation w.r.t. $g_{\alpha\beta}$ and the indices of
$K^{\mu_1\cdots\mu_m}$ are lowered using $g_{\alpha\beta}$. This is
the Killing tensor equation and a solution $K_{\mu_1\cdots\mu_m}$ is
known as a Killing tensor of rank $m$. A closely 
related geometric object is a conformal Killing tensor, a symmetric tensor
$P_{\mu_1\cdots\mu_m}$ for which $P^{\mu_1\cdots\mu_m}p_{\mu_1} \cdots
p_{\mu_m}$ in general is invariant for null geodesics only. Since this
is nontrivial only for the trace-free part of any tensor, a conformal
Killing tensor is usually taken to be trace-free from the outset. The
conformal Killing tensor equation then takes the form 
\begin{equation}\lbeq{ckteqm}
  P_{(\mu_1\cdots\mu_m;\mu_{m+1})} =
  m\,[2(m-1)+n]^{-1}g_{(\mu_1\mu_2}P^\nu{}_{\mu_3\cdots\mu_{m+1});\nu},
\end{equation}
where $n$ is the dimension of the configuration space. In particular,
the trace-free part of any Killing tensor is a conformal Killing
tensor satisfying this equation.

In this work we focus on the classical two-dimensional Hamiltonians
which can be written as 
\begin{equation}
  H = \textstyle\frac12(p_x{}^2 + p_y{}^2) + V(x,y).
\end{equation}
The associated Jacobi geometry is given by the line element 
\begin{equation}
  ds^2 = 2G(dx^2+dy^2) = 2Gdzd\bar{z}, \quad G=E-V,
\end{equation}
where we have introduced the complex conjugate null variables
$z=x+iy$, $\bar{z}=x-iy$, which are convenient to use as they are
adapted to the action of the conformal group. This will make the
conformal Killing tensor equation \refeq{ckteqm} maximally
simplified. Furthermore, all tensor calculations will be done in the
standard null frame 
$\Omega^0 = G^{1/2}dz$, $\Omega^{\bar{0}} = G^{1/2}d\bar{z}$ in which
the metric takes the simplest possible form $ds^2=2d\Omega^0
d\Omega^{\bar{0}}$. 
We use the convention that tensor indices in this frame take the
values $0$ and $\bar{0}$, while in any coordinate frame the values
will be the names of the coordinates (e.g.\ $z$ and $\bar{z}$).

\section{Mapping invariants between the physical and the Jacobi time
  gauge}\label{section:mapping}

We consider a Hamiltonian of the type \refeq{hamclass},
admitting a cubic invariant $I$ at
least on some fixed energy surface $H=E$ and possibly at arbitrary
energy. The invariant can without restriction be assumed to have the
form
\begin{equation}\lbeq{I3phys}
  I = A^{\alpha\beta\gamma}p_\alpha p_\beta p_\gamma + B^\alpha
  p_\alpha,
\end{equation}
since an additional term that is even in the momenta must Poisson
commute separately with $H$. To see how the invariant
transforms when going to the Jacobi time gauge, we perform the
time reparametrization by using the method of coupling constant
metamorphosis \cite{hietarinta:ccm}. To this end we wish to introduce
a coupling constant $\kappa$ into the 
Hamiltonian by rescaling the potential according to $V\rarr 2\kappa
V$. To see how this rescaling affects the invariant, it
is useful to implement it in two steps, which both trivially preserve 
Poisson commutativity although they are in fact noncanonical
transformations. The first step is to rescale
the momenta according to $p_\alpha \rarr \lambda^{-1}p_\alpha$ which
gives 
\begin{equation}
\begin{split}
  H &\rarr \lambda^{-2}T+V, \\
  I &\rarr \lambda^{-3}A^{\alpha\beta\gamma}p_\alpha p_\beta p_\gamma + 
  \lambda^{-1} B^\alpha p_\alpha, \\
  \{\;,\:\} &\rarr \lambda \{\;,\:\},
\end{split}
\end{equation}
where $\{\;,\:\}$ is the Poisson bracket. By the bilinearity of the
Poisson bracket, we are also free to rescale the commuting functions
$H$ and $I$ themselves. The second step is to set
$\kappa=\frac12\lambda^2$ and use this rescaling freedom to redefine
$H$ and $I$ according to
\begin{equation}
\begin{align}\lbeq{Hredef}
  \tilde{H} &:= \lambda^2 H = T + 2\kappa V, \\ \lbeq{Iredef}
  \tilde{I} &:= \lambda^3 I = A^{\alpha\beta\gamma} p_\alpha p_\beta
  p_\gamma + 2\kappa B^{\alpha}p_\alpha,  
\end{align}
\end{equation}
which gives the desired rescaling of $V$. We could of course have
defined $\tilde{I}$ using any scaling factor, but $\lambda^3$ will
turn out to be the natural general choice.   
As in \cite{goliath-et-al:laxarb}, to obtain the Jacobi Hamiltonian
when solving a fixed energy constraint for $\kappa$, we must hold on
to the interpretation of the parameter $E$ as the energy value of the
original Hamiltonian $H=T+V$, thus making 
$2\kappa E$ the corresponding energy value of the rescaled Hamiltonian
$\tilde{H}$. Solving $\tilde{H}=2\kappa E$ for $\kappa$ then
results in 
\begin{equation}
  \kappa = [2(E-V)]^{-1}T = H_J,
\end{equation}
where we have used the standard definition of the Jacobi Hamiltonian
$H_J$ associated with the fixed energy surface of the original
Hamiltonian $H=T+V$.  
Thus we find that a coupling constant metamorphosis acting on $\kappa$
is equivalent to a transformation to Jacobi time. However, as noted in
\cite{goliath-et-al:laxarb}, this is not a coupling constant
metamorphosis in the true sense, as the old energy $E$ does not
enter linearly into the new Hamiltonian $H_J$. Nevertheless the
results of \cite{hietarinta:ccm} still apply which means that the
physical invariant $I$ and Jacobi invariant $I_J$, Poisson commuting
with $H$ and $H_J$ respectively, are transformed into 
each other according to
\begin{equation}\lbeq{recipe}
  I_J = \left.\tilde{I}\right|_{\kappa\rarr H_J}, \quad I =
  \left.I_J\right|_{E\rarr H}.
\end{equation}
Hence the Jacobi invariant corresponding to the physical invariant
\refeq{I3phys} is 
\begin{equation}\lbeq{IJ3}
  I_J = A^{\alpha\beta\gamma}p_\alpha
  p_\beta p_\gamma + 2H_J B^\alpha p_\alpha =
  K^{\alpha\beta\gamma}p_\alpha p_\beta p_\gamma, \quad
  K^{\alpha\beta\gamma} := 
A^{\alpha\beta\gamma}+B^{(\alpha}g^{\beta\gamma)},
\end{equation}
where $g^{\alpha\beta}$ is the inverse of the Jacobi metric
$g_{\alpha\beta} := 2(E-V)h_{\alpha\beta}$. It follows directly that
$K_{\alpha\beta\gamma}$ is a third rank Killing tensor w.r.t. the
Jacobi metric. Thus we see that a cubic invariant of the form
\refeq{I3phys} can always be mapped to such a Killing tensor. However,
there might in fact be cases where the invariant \refeq{IJ3} is
reducible and for which a nontrivial cubic invariant more naturally
corresponds to a Killing vector, rather then a third rank Killing
tensor. By appyling the transformation recipe \refeq{recipe}, one
realizes that for this to happen, the cubic invariant \refeq{I3phys}
must be of the form
\begin{equation}\lbeq{I3physspec}
  I = H C^\alpha p_\alpha + D^\alpha p_\alpha,
\end{equation}
for some vectors $C^\alpha$, $D^\alpha$. This is clearly the same as
saying that the tensor $A^{\alpha\beta\gamma}$ has a vanishing
traceless part. Now, if after taking $p_\alpha\rarr
\lambda^{-1}p_\alpha$ we would have defined the rescaled invariant
\refeq{I3physspec} according to $\tilde{I} := \lambda I$ instead of
$\tilde{I} \rarr \lambda^3I$ as was done in eq.\ \refeq{Iredef}, the 
corresponding Jacobi invariant would simply become 
\begin{equation}\lbeq{kvcub}
  I_J = \xi^\alpha p_\alpha, \quad \xi^\alpha = EC^\alpha+D^\alpha
\end{equation} 
from which it follows directly that $\xi_\alpha$ is a Killing vector 
w.r.t. the Jacobi metric. In section \ref{section:killingvector} we
give an  example of a Hamiltonian which has a nontrivial cubic
invariant of the type \refeq{I3physspec}.

\section{Cubic invariants corresponding to third rank Killing tensors} 

In this section we begin by deriving a necessary and sufficient
integrability condition for the Jacobi metric to admit a third rank
Killing tensor at a fixed value of the energy parameter $E$. In the
next section we proceed by finding the conditions that ensure that the
Killing tensor equations will be satisfied at arbitrary energy values,
but with attention restricted to the case when the energy dependence
of the Killing tensor is such that the corresponding invariant of the
physical Hamiltonian is cubic. In analogy with the second rank case
\cite{rp:inv} as well as the 
third rank case with indefinite metric \cite{mk-kr:1+1-rank3}, our
Killing tensor $K_{\alpha\beta\gamma}$
will from the outset be decomposed into its trace-free (conformal) part
$P_{\alpha\beta\gamma}$ and trace $K_\alpha := K^\beta{}_{\beta\alpha}$
according to
\begin{equation}
  K_{\alpha\beta\gamma} = P_{\alpha\beta\gamma} + 3\,(n+2)^{-1}
  K_{(\alpha}g_{\beta\gamma)},
\end{equation}
where $n$ is the dimension of the configuration space, i.e. $n=2$ in
our case. Similarly the Killing tensor equation itself,
$K_{(\alpha\beta\gamma;\delta)} = 0$, will be split into its three
trace-free ``components'', namely its trace-free part, the trace-free
part of its trace and the trace of its trace, which read
\begin{align}\lbeq{ckteq}
    &C_{\alpha\beta\gamma\delta} := P_{(\alpha\beta\gamma;\delta)} - 
    3\,(n+4)^{-1}g_{(\alpha\beta}P^\lambda{}_{\gamma\delta);\lambda} = 0
    \\ \lbeq{couplingeq}
    &D_{\alpha\beta} := K_{(\alpha;\beta)} +
    (n+2)\,(n+4)^{-1}P^\gamma{}_{\alpha\beta;\gamma} = 0 \\
    \lbeq{divfreeeq} 
    &K^\alpha{}_{;\alpha} = 0, 
\end{align}
where the trace of eq.\ \refeq{couplingeq} is automatically satisfied
once the divergence-free condition \refeq{divfreeeq} for $K_\alpha$
has been solved. Just as in the case of an indefinite metric
\cite{mk-kr:1+1-rank3}, it is advantageous to use the coordinate frame
components of $K_\alpha$ and $P^{\alpha\beta\gamma}$ (note the index
positioning) when parametrizing the four independent components of
$K_{\alpha\beta\gamma}$. The parametrization thus becomes
\begin{equation}
\begin{split}
  K_{000}                   &= G^{3/2}P^{\bar{z}\bar{z}\bar{z}} \\
  K_{\bar{0}\bar{0}\bar{0}} &= G^{3/2}P^{zzz} \\
  K_{00\bar{0}}             &= \textstyle \frac12G^{-1/2}K_z \\
  K_{0\bar{0}\bar{0}}       &= \textstyle \frac12G^{-1/2}K_{\bar{z}}
\end{split} 
\end{equation}
The calculations leading to the integrability condition imposed on the
Jacobi metric are analogous to the indefinite case
\cite{mk-kr:1+1-rank3}. An important difference, however, is the fact
that the null variables $z$ and $\bar{z}$ here are complex conjugate.
Demanding that the potential $V$ (and thereby the Jacobi metric) be
real implies that there is no restriction in assuming that the Killing
tensor $K_{\alpha\beta\gamma}$ be real as well. This in turn leads to
the component constraints that $K_{\bar{z}}$ and
$P^{\bar{z}\bar{z}\bar{z}}$ be the complex conjugates of $K_z$ and
$P^{zzz}$, respectively. In the indefinite case, on the other hand,
the reality condition for the Killing tensor requires that the
corresponding four components be real, but implies no other relation
between them.

We now proceed by solving the two equations \refeq{ckteq} and
\refeq{divfreeeq}, leaving \refeq{couplingeq} - the only equation which 
couples the conformal and trace parts of the Killing tensor - for
later consideration. 

The conformal Killing tensor equation \refeq{ckteq} for
$P_{\alpha\beta\gamma}$ has the components 
\begin{equation}
  \overline{C_{0000}} = C_{\bar{0}\bar{0}\bar{0}\bar{0}} =
    GP^{zzz}{}_{,\bar{z}} = 0.
\end{equation}
Hence $P_{\alpha\beta\gamma}$ can be fully represented in terms of an
analytic function $S(z) := P^{zzz}$. We shall assume $S(z)\neq 0$
to be the case, since otherwise eq.\ \refeq{couplingeq} implies that
$K_\alpha$ is a Killing vector. The divergence free condition
\refeq{divfreeeq} for $K_\alpha$,
\begin{equation}
  K^{\alpha}{}_{;\alpha} = 2G^{-1}\real\{K_{z,\bar{z}}\} = 0,
\end{equation}
is just as easily solved by introducing a real potential function
$\Phi$ satisfying $K_z = 2i\Phi_{,z}$, with the factor $2$ inserted
for later convenience. This potential relation can also 
be expressed covariantly in terms of the natural volume $2$-form
$\epsilon_{\alpha\beta} =
i(\Omega^0\wedge\Omega^{\bar{0}})_{\alpha\beta}$\, as
\begin{equation}
  K_{\alpha} = 2\epsilon_\alpha{}^\beta\Phi_{;\beta},
\end{equation}
provided that $\Phi$ transforms as a scalar. We can now write eq.\ 
\refeq{couplingeq} in the form
\begin{equation}
  D_{\alpha\beta} = 2\Phi_{;\gamma(\alpha}\epsilon_{\beta)}{}^\gamma + 
  \textstyle\frac23P^\gamma{}_{\alpha\beta;\gamma} = 0.
\end{equation}
Its components,
\begin{equation}\lbeq{couplingcomp}
  \overline{D_{00}} = D_{\bar{0}\bar{0}} =
  -2i(\frac{\Phi_{,\bar{z}}}{G})_{,\bar{z}}
  +\textstyle\frac23G^{-2}(G^3S)_{,z} = 0,
\end{equation}
can in analogy with the procedure used in \cite{mk-kr:1+1-rank3} be
simplified by using a conformal transformation to a new complex null
variable $w = H(z)$, with the inverse relation being $z = F(w)$. The
metric will then be given in terms of the new conformal factor
$\tilde{G}=|F'(w)|^2G$ as $ds^2=2\tilde{G}dw d\bar{w}$. To preserve
the relation between the standard null frame and the null variable,
the frame must be scaled as $\tilde{\Omega}^0 = B\Omega^0$,
$\tilde{\Omega}^{\bar 0} = B^{-1}\Omega^{\bar{0}}$ where $B^{-1} =
\bar{B} = [F'(w)/\bar{F}(\bar{w})]^{1/2}$. The analytic function
$S(z)$ transforms as
\begin{equation}\lbeq{Stransf}
  \tilde{S}(\omega) := P^{www} = [H'(z)]^3P^{zzz} =
  [H'(z)]^3S(z). 
\end{equation}
This shows that we can always make $\tilde{S}(w)$ take the standard
constant value $1$ by choosing the conformal transformation such that
$H'(z)=[S(z)]^{-1/3}$. With this choice, eq.\ \refeq{couplingcomp}
simplifies to 
\begin{equation}\lbeq{couplingsimp}
  \overline{\tilde{D}_{00}} = \tilde{D}_{\bar{0}\bar{0}} =
  -2i(\frac{\Phi_{,\bar{w}}}{\tilde{G}})_{,\bar{w}} 
  + 2\tilde{G}_{,w} = 0.
\end{equation}
Comparing this equation with its complex conjugate leads directly
to the integrability condition
\begin{equation}
  \real\{(\frac{\Phi_{,w}}{\tilde{G}})_{,ww}\} = 0, 
\end{equation}
which is satisfied iff $\Phi$ and $\tilde{G}$ are related by some real
function $\mcl{K}$ according to 
\begin{equation}\lbeq{Kintro}
  \Phi_{,w} = i\tilde{G}\mcl{K}_{,\bar{w}\bar{w}}.
\end{equation}
Substituting this back into eq.\ \refeq{couplingsimp} yields
\begin{equation}
  \tilde{G}_{,w} = \mcl{K}_{,ww\bar{w}}.
\end{equation}
In fact, as $\mcl{K}$ is determined only up to the
transformation $\mcl{K} \rarr \mcl{K} + Aw\bar{w} +
\real\{\Lambda(w)\}$, where $A$ is a real constant and
$\Lambda(w)$ is an analytic function, we can partially fix this 
freedom by requiring that the relation 
\begin{equation}\lbeq{kahler}
  \tilde{G} = \mcl{K}_{,w\bar{w}}
\end{equation}
holds, meaning that $\mcl{K}$ becomes a K\"ahler potential
\cite{nakahara} for the Jacobi metric. As such, $\mcl{K}$ transforms
as a scalar under conformal transformations. Hence for \emph{any} null
variable $z$, the metric conformal factor $G$ is related to $\mcl{K}$
according to the simple formula $\mcl{K}_{,z\bar{z}} = G$.
Substituting eq.\ \refeq{kahler} into eq.\ \refeq{Kintro} now leads to
the final integrability condition
\begin{equation}\lbeq{Kcondstand}
  \real\{
  (\mcl{K}_{,w\bar{w}}\mcl{K}_{,ww})_{,w} \}
  = 0,   
\end{equation}
which is necessary and sufficient for the existence of a third rank
Killing tensor. In contrast to the first and
second rank cases \cite{rp:inv}, we are here dealing with a nonlinear
condition imposed on the metric. It can be noted that a condition
which is formally identical to eq.\ \refeq{Kcondstand} also applies to
one of the three different types of third rank Killing tensors that
are allowed when the metric has indefinite signature
\cite{mk-kr:1+1-rank3}. The interested reader can consult that
reference for a number of nontrivial solutions. Moreover, the same
condition can be found in \cite{hietarinta:secinv}, but not in the
context of the geometric approach of this work in which the unknown
function can be interpreted as a K\"{a}hler potential.

Going back to the original null variable $z$, eq.\ \refeq{Kintro} now
takes the form
\begin{equation}\lbeq{PhiK}
  \Phi_{,z} = i\mcl{K}_{,z\bar{z}}
  (\bar{S}\mcl{K}_{,\bar{z}\bar{z}} +
  \textstyle\frac13\bar{S}'\mcl{K}_{,\bar{z}}),   
\end{equation}
with the corresponding integrability condition
\begin{equation}\lbeq{Kintgen}
  \textstyle
  \real\{[\mcl{K}_{,z\bar{z}}(S\mcl{K}_{,zz}+\frac13S'\mcl{K}_{,z})]_{,z}\} 
  = 0.
\end{equation}
The cubic Jacobi invariant $I_J = K^{\alpha\beta\gamma}p_\alpha
p_\beta p_\gamma$ can now, via eq.\ \refeq{PhiK} and
$G=\mcl{K}_{,z\bar{z}}$, be written down as
\begin{equation}\lbeq{IJgen}
  I_J = 2\real\{Sp_z{}^3 -
  (3S\mcl{K}_{,zz} +
  S'\mcl{K}_{,z})H_J p_z\}, \quad H_J = \mcl{K}_{,z\bar{z}}{}^{-1}p_z
  p_{\bar{z}}. 
\end{equation}
Expressed in the standardized null variable $w$, this simplifies to 
\begin{equation}
  I_J = 2\real\{p_w{}^3 -
  3\mcl{K}_{,ww}H_J p_w\}, \quad H_J =
  \mcl{K}_{,w{\bar{w}}}{}^{-1} p_w p_{\bar{w}}. 
\end{equation}
Note that up to conformal transformations, the Jacobi Hamiltonian and
its cubic invariant is completely determined by $\mcl{K}$.

\section{Arbitrary energy invariants}

So far we have derived the necessary and sufficient condition
(eq.\ \refeq{Kcondstand} or eq.\ \refeq{Kintgen}) for a
two-dimensional Riemannian geometry to admit a third rank Killing
tensor. Given that this condition is satisfied, we can use any null
variable $z$ and identify the associated metric conformal factor
$G=\mcl{K}_{,z\bar{z}}$ with $E-V$ for a flat space Hamiltonian 
$H=2p_zp_{\bar{z}}+V$ which is integrable with cubic invariant at
fixed energy $E$. Thus we can choose to interpret the geometry as a
common Jacobi geometry of a large family of such flat space
Hamiltonians, whose members are related by conformal
transformations. Now, however, we 
shall make the identification $\mcl{K}_{,z\bar{z}} = E-V$ for a
particular null variable $z$ and  
derive the additional conditions that make the Killing tensor
equations satisfied for all values of $E$, corresponding to
$H=2p_zp_{\bar{z}}+V$ being integrable at arbitrary energy. To this 
end we begin by noting that we can write the K\"ahler
potential as
\begin{equation}
  \mcl{K} = Ez\bar{z}+2\real\{\theta(E,z)\}-\Psi,
\end{equation} 
where $\theta(E,z)$ is analytic in $z$ and can be assumed to satisfy
$\theta(0,z)=0$,
whereas $\Psi$ is real, satisfies $\Psi_{,z\bar{z}}=V$ and is
independent of $E$. It appears at first sight that $\theta(E,z)$ in
principle could have any dependence on $E$, which would make it very
difficult to 
continue working in full generality. Similarly it seems
possible that the function 
$S(z)$ could also have some dependence on $E$. However, if we apply
the recipe that the Jacobi invariant $I_J$ is transformed to the
physical time gauge 
according to $I = \left.I_J\right|_{E\rarr H}$, it follows from eq.\
\refeq{Kintgen} and eq.\ \refeq{IJgen} that $I$ becomes a cubic
polynomial invariant 
iff $S(z)$ is  independent of $E$ and no restriction is implied by
setting $\theta(E,z) = E\Lambda(z)$ for some analytic function
$\Lambda(z)$ which is also independent of $E$. Since this paper is
devoted to cubic invariants, we shall use this fact from the
outset. Our physical cubic invariant $I$ then has the general form 
\begin{equation}\lbeq{cubicI}
  I = \real \{2Sp_z{}^3 - 
    2[S'(\bar{z}+\Lambda')+3S\Lambda'']p_z{}^2p_{\bar{z}}
    +\textstyle[S'(\Psi_{,z} -
    (\bar{z}+\Lambda')\Psi_{,z\bar{z}}) + 
    3S(\Psi_{,zz} - \Lambda''\Psi_{,z\bar{z}})]p_z\}.
  \end{equation}
If we now substitute
\begin{equation}\lbeq{KarbE}
  \mcl{K} = E[z\bar{z}+2\real\{\Lambda(z)\}] - \Psi
\end{equation}
into eq.\ \refeq{Kintgen}, the condition becomes a second degree
polynomial in $E$: 
\begin{equation}
  A_2E^2+A_1E+A_0 = 0,
\end{equation}
where the coefficients $A_i$ must vanish separately if the equation is
to hold for arbitrary $E$. This gives the three equations 
\begin{equation}
\begin{align}\lbeq{A2}
  A_2 &= \real\left\{[S'(\bar{z}+\Lambda')+3S\Lambda'']_{,z}\right\} =
  0, \\ \lbeq{A1} 
  A_1 &= -\real\left\{[S'((\bar{z}+\Lambda')\Psi_{,z\bar{z}}+\Psi_{,z}) +
  3S(\Lambda''\Psi_{,z\bar{z}}+\Psi_{,zz})]_{,z}\right\} = 0, \\ \lbeq{A0}
  A_0 &=
  \real\left\{[\Psi_{,z\bar{z}}(S'\Psi_{,z}+3S\Psi_{,zz})]_{,z}\right\}
  = 0. 
\end{align}
\end{equation}
By applying the differential operator $\partial^2/ \partial z
  \partial \bar{z}$ to eq.\ \refeq{A2}, we split off
the condition
\begin{equation}
  \real\{S'''(z)\} = 0,
\end{equation}
with the polynomial solution
\begin{equation}\lbeq{Spoly}
  S(z) = iaz^3 + \beta z^2 + \gamma z + \delta,
\end{equation}
where $a$ is a real constant and $\beta$, $\gamma$ and
$\delta$ are complex constants. A standardization of these
coefficients can be achieved by using the transformation property
of $S(z)$ given by eq.\ \refeq{Stransf}, combined with the freedom
to make translations $z\rarr z+z_0$, rotations $z\rarr e^{ib}z$ and
scalings $z\rarr cz$, with $b$ and $c$ real. Comparing with the first
and second rank cases \cite{rp:inv} leads us to conjecture that the
analytic function $S(z)$ representing the conformal part of a 
Killing tensor of arbitrary rank $m$ is required to satisfy 
\begin{equation}
\begin{split}
  \real\{(\textstyle\frac d{dz})^mS(z)\} &= 0 \quad
  (m\:\mathrm{odd}),  \\ 
  \imag\{(\textstyle\frac d{dz})^mS(z)\} &= 0 \quad
  (m\:\mathrm{even}). \\   
\end{split}
\end{equation}
However, we make no attempt to prove this. For any choice of
coefficients $a$, $\beta$, $\gamma$ and $\delta$, eq.\ \refeq{A2} is a 
linear ODE in $\Lambda(z)$ which can be solved by standard
methods. Once this first equation has been taken care of, eq.\
\refeq{A1} and \refeq{A0} are two PDEs for one unknown function
$\Psi$, i.e. in general an overdetermined system. This is consistent
with the fact that only isolated cases of classical Hamiltonians
integrable with cubic invariant are known
\cite{hietarinta:secinv}. The obvious way to proceed is to try to find
the general solution to the linear PDE \refeq{A1} and then check if
its functional degrees of freedom can be exploited to make the
nonlinear PDE \refeq{A0} satisfied. As a simple illustration of this
method, we start out from the simplest possible solution to
eq.\ \refeq{A2}, namely $a=\beta=\gamma=\Lambda(z)=0$, $\delta=d$ with
$d$ real. In terms of the real variables $x$ and $y$ which satisfy
$z=x+iy$, the linear condition \refeq{A1} in this case becomes
\begin{equation}
  \Psi_{,xxx} - 3\Psi_{,xyy} = 0.
\end{equation}
Writing down the general solution as
\begin{equation}
  \Psi = f_1(y+\sqrt3\,x) + f_2(y-\sqrt3\,x) + f_3(-2y), 
\end{equation}
the nonlinear condition \refeq{A0} takes the cyclic form 
\begin{equation}\lbeq{laxcond}
  F_1{}'(F_2-F_3) + F_2{}'(F_3-F_1) + F_3{}'(F_1-F_2) = 0,
\end{equation}
where $F_i = f_i{}''$ so that 
\begin{equation}\lbeq{lax2dim}
  V = F_1(y+\sqrt3\,x) + F_2(y-\sqrt3\,x) + F_3(-2y). 
\end{equation}
The functional condition \refeq{laxcond} is well-known from
Lax pair studies \cite{hietarinta:secinv,os-per:classint} and arises
for a three-particle Hamiltonian of the generic form 
\begin{equation}
  \textstyle H = \frac12(p_1{}^2 + p_2{}^2 + p_3{}^2) + F_1(q^3-q^2) +
  F_2(q^1-q^3) + F_3(q^2-q^1),
\end{equation}
which can be reduced to our two-dimensional Hamiltonian with $V$ given 
by eq.\ \refeq{lax2dim} by going to the center-of-mass system and
making a rescaling of the coordinates \cite{hietarinta:secinv}. 
Its solutions include the Toda potential $F_i(\xi) =
C_ie^{b\xi}$ and also $F_i(\xi) = C\mcl{P}(b\xi)$, where the
Weierstrass function $\mcl{P}(\xi)$ can be taken in the special limits 
$\mcl{P}(\xi)\rarr\xi^{-2}$, $\mcl{P}(\xi)\rarr\sinh^{-2}\xi$ and
$\mcl{P}(\xi)\rarr\sin^{-2}\xi$. The cubic invariant of a system of
this type can be found in table \ref{table:S0}. 
 
In more general cases, the linear PDE \refeq{A1} is not as easily
solved. Failing to work in full generality, we have restricted
attention to the special cases where $S(z)$ is a homogeneous
polynomial, i.e. we have set all but one of the constants $a$,
$\beta$, $\gamma$ and $\delta$ to zero. Furthermore, for all our
solutions, whether previously known or not, it has turned out that
$\Lambda(z)=\lambda z^2$ for some complex constant $\lambda$. Thus the
coefficient $z\bar{z}+2\real\{\Lambda(z)\}$ of $E$ in the K\"ahler
potential \refeq{KarbE} is always a hermitian form in $z$ (or
equivalently, a real quadratic form in $x$ and $y$). However, we
stress that solutions of other types may very well exist. With both
$S(z)$ and $z\bar{z}+2\real\{\Lambda(z)\}$ being homogeneous functions
of $z$, $\bar{z}$, it is very natural to make the anzats that $\Psi$
is homogeneous as well. To this end one can e.g. introduce polar
coordinates $r$, $\phi$ according to $z=re^{i\phi}$ and set $\Psi =
r^k f(\phi)$, where $k$ is a real constant and $f(\phi)$ is an
arbitrary real function. This form of $\Psi$ gives a Jacobi metric 
\begin{equation}
  ds^2=2(E-V)(dr^2+r^2d\phi^2), \quad V = \textstyle\frac14
  r^{k-2}[k^2f(\phi)+f''(\phi)]
\end{equation}
which, at zero energy (all energies iff $k=2$), admits a homothetic
vector field $\zeta = 2k^{-1}r\,\partial/\partial r$, satisfying
$\mathcal{L}_\zeta g_{\alpha\beta} = 2g_{\alpha\beta}$. The advantage
of this ansatz is that eq.\ \refeq{A1} and 
\refeq{A0} become ODEs in $f(\phi)$, parametrized by $k$. Several
of our solutions, some of which are new, were found using this   
approach. However, it is worth mentioning that in some cases the ODEs become 
simpler in other coordinates (such as $\xi=z$, $\eta=\bar{z}/z$) which
are also adapted to the homogeneous ansatz.

\section{Cubic invariants corresponding to Killing
  vectors}\label{section:killingvector} 

In this section we look for geometries admitting a Killing vector
$\xi_\alpha$ corresponding to a cubic invariant by the mechanism
discussed in section \ref{section:mapping}. If we parametrize the
Killing vector by $\xi^z$ and its complex conjugate, the trace-free 
part of the Killing vector equations $\xi_{(\alpha;\beta)}=0$
(i.e. the conformal Killing vector equations) becomes
\begin{equation}
  \overline{\xi_{(0;0)}} = \xi_{(\bar{0};\bar{0})} =
  \xi^z{}_{,\bar{z}} = 0,
\end{equation}
which is solved by setting $\xi^z=Z(z)$ with $Z(z)$ analytic. The
linear Jacobi invariant thus takes the form $I_J = \xi^\alpha p_\alpha 
= 2\real\{Z(z)p_z\}$. The trace part of the Killing vector equations
becomes 
\begin{equation}\lbeq{kveqtrace}
  \xi^{\alpha}{}_{;\alpha} = 2G^{-1}\real\{(SG)_{,z}\} = 0.
\end{equation}
Under conformal transformations, $S(z)$ transforms according to
\begin{equation}\lbeq{Ztransf}
  \tilde{Z}(w) := \xi^w = H'(z)\xi^z = H'(z)Z(z).
\end{equation}
Hence we obtain $\tilde{Z}(w)=1$ by choosing the 
transformation so that $H'(z)=[Z(z)]^{-1}$, which makes eq.\
\refeq{kveqtrace} take the standardized form
\begin{equation}\lbeq{kvstand}
  \real\{\tilde{G}_{,w}\} = 0.
\end{equation}
It follows that $\tilde{G}$ is a function of
$Y=\imag w$ only, which is consistent with the invariant being $I_J =
2p_X$, $X = \real w$. 

We now fix a null variable $z$ for which we set $G=E-V$ and focus on
trying to make the Killing vector 
equations hold for arbitrary energy $E$, for the case that the Killing
vector corresponds  to a cubic physical invariant. According to eq.\
\refeq{kvcub} we must allow for $Z(z)$ to have a linear dependence on
$E$: 
\begin{equation}\lbeq{ZlinE}
  Z(z) \rarr Z(E,z) = Z_1(z)E + Z_0(z),
\end{equation}
where we have set $Z_1(z) = C^z$ and $Z_0(z) = D^z$.
This leads to the physical invariant taking the general form
\begin{equation}\lbeq{kvcubinv}
  I = 2\real\{(Z_1H+Z_0)p_z\}, \quad H = 2p_zp_{\bar{z}} + V.
\end{equation}
Substituting eq.\ \refeq{ZlinE} into eq.\ \refeq{kveqtrace} yields
\begin{equation}
  \textstyle\frac12\real\{(ZG)_{,z}\} = B_2E^2 + B_1E + B_0 = 0,
\end{equation}
giving us the three equations
\begin{equation}
\begin{align}\lbeq{B2}
  B_2 &= \real\{Z_1{}'\} = 0, \\ \lbeq{B1} 
  B_1 &= \real\{(Z_0-Z_1V)_{,z}\} = 0, \\ \lbeq{B0}
  B_0 &= -\real\{(Z_0V)_{,z}\} = 0. 
\end{align}
\end{equation}
The solution to eq.\ \refeq{B2} reads
\begin{equation}
  Z_1(z) = iaz + \beta,
\end{equation}
where $a$ is a real constant and $\beta$ is a complex constant. This
gives us two inequivalent cases to study since we can make a linear
transformation such that either $Z_1(z)\rarr 1$ or $Z_1(z)\rarr iz$
depending on whether $a$ is zero or nonzero. We consider these two
cases in some detail below.

\subsubsection*{The Case $Z_1(z)=1$}

In this case eq. \refeq{B1} takes the form
\begin{equation}
  \real\{ Z_0{}'-V_{,z} \} = 0,
\end{equation}
with the general solution given by
\begin{equation}
  V = Z_0(z)+\bar{Z}_0(\bar{z})+F(y), \quad y=\imag z,
\end{equation}
Here $F(y)$ is an arbitrary real function. The final condition
\refeq{B0} now becomes the functional equation
\begin{equation}
  \textstyle\real \{ [Z_0(Z_0+\bar{Z}_0+F)]_{,z} \} =
  (Z_0+\frac12\bar{Z}_0+\frac12F)Z_0{}' +
  (\bar{Z}_0+\frac12Z_0+\frac12F)\bar{Z}_0{}' -
  \frac{i}4(Z_0-\bar{Z}_0)F' = 0. 
\end{equation}
We have found no nontrivial solution to this equation.

\subsubsection*{The Case $Z_1(z)=iz$}

Here eq.\ \refeq{A1} reads
\begin{equation}
  \real\{ Z_0{}'-izV_{,z} \} = 0,
\end{equation}
and is solved by
\begin{equation}
  V = i\left(-\int\!\frac{Z_0{}'}{z}dz +
  \int\!\frac{\bar{Z}_0{}'}{\bar{z}}d\bar{z}\right) + F(r),
\quad r = \sqrt{z\bar{z}}, 
\end{equation}
with $F(r)$ being an arbitrary real function. The integrals can
be avoided by introducing an analytic function $Q(z)$ satisfying 
\begin{equation}
  Z_0 = i(zQ'-Q).
\end{equation}
We can then write the potential as
\begin{equation}
  V = Q'(z)+\bar{Q}'(\bar{z})+F(r),
\end{equation}
which makes eq.\ \refeq{B0} become
\begin{equation}
\begin{split}
  \imag\{ [(zQ'-Q)(Q'+\bar{Q}'+F)]_{,z} \} &= \ts
  \frac{1}{2i}(2zQ'-Q+z\bar{Q}'+zF)Q'' -
  \frac{1}{2i}(2\bar{z}\bar{Q}'-\bar{Q}+\bar{z}Q'+\bar{z}F)\bar{Q}''
  \\  
  &+ \ts\frac{1}{4i}(rQ'-r\bar{Q}'-\bar{z}Q/r+z\bar{Q}/r)F' = 0.
\end{split}
\end{equation}
This functional equation has at least one nontrivial solution, given
by $Q(z) = \alpha \sqrt{z}$, $F(r) = Ar^{-1}$ with $\alpha$ complex
and $A$ real. Writing $\alpha = \sqrt2(B+iC)$, the potential becomes
\begin{equation}\lbeq{kvcubV}
  V = \frac{A + B\sqrt{r+x}+C\sqrt{r-x}}{r}.
\end{equation}
This potential was given in complex form by Drach
\cite{drach:invcubic}. We also note that the potential also admits a
quadratic invariant \cite{hietarinta:secinv} and is therefore now
superintegrable. The cubic invariant of the system, given by eq.\ 
\refeq{kvcubinv}, is presented in table \ref{table:kvinv}. What is
interesting about this result is that it gives a new way of solving
Hamilton's equations for the system, even in the physical time gauge.
Namely, for any fixed energy $E$, one can choose coordinates such that
the cubic invariant $I$ reduces to the momenta $p_Q$ of a
configuration coordinate $Q$ that becomes cyclic in the Hamiltonian
for this particular energy value. In other words the equations of
motion can be solved just as in the case of an ordinary linear
invariant, but the coordinate transformation which makes $Q$ cyclic is
dependent on the energy $E$. In the geometric picture, one obtains a
cyclic variable by the above described conformal transformation which
makes the metric conformal factor satisfy eq.\ \refeq{kvstand}.

Quite surprisingly, we soon realized that this potential via coupling 
constant metamorphosis acting on $A$ is dual to the usual (isotropic)
harmonic oscillator potential, with its center in general translated
off the coordinate origin. This coupling constant metamorphosis can be
realized as a conformal transformation in the Jacobi geometry
setting. Choosing the transformation as $z=F(w)=w^2$, the metric
conformal factor $G=E-V$ with $V$ given by eq.\ \refeq{kvcubV},
transforms into
\begin{equation}
\begin{split}
  \tilde{G} = |F'|^2 G &= \tilde{E} - \tilde{V}, \quad \tilde{V} =
  \tilde{A}(X^2+Y^2) + \tilde{B}X + \tilde{C}Y. 
\end{split}
\end{equation}
where $X=\real{w}$, $Y=\imag{w}$, $\tilde{E}=-4A$, $\tilde{A}=-4E$,
$\tilde{B}=4\sqrt2 B $ and
$\tilde{C}=4\sqrt2{C}$. We have assumed that $X$ and $Y$ are both
positive to avoid keeping track of signs. It follows that $\tilde{V}$
can be interpreted as a potential which is integrable at arbitrary energy
$\tilde{E}$. Obviously this $\tilde{V}$ is just the translated
harmonic oscillator potential.

The analytic function $Z(E,z) = i[Ez-2^{-1/2}(B+iC)\sqrt{z}]$, on the
other hand, transforms according to eq.\ \refeq{Ztransf} into   
\begin{equation}
  \textstyle\tilde{Z}(w) = -\frac{i}{16}(2\tilde A w + \tilde B +
  i \tilde C).
\end{equation}
Since $\tilde{Z}(w)$ is independent of the new energy $\tilde{E}$, the
invariant is the same in both time gauges and reads
\begin{equation}
  I = I_J = \textstyle\frac1{16}
  [(2\tilde{A}Y+\tilde{C})p_X-(2\tilde{A}X+\tilde{B})p_Y],  
\end{equation}
where, of course, the factor $\frac1{16}$ could be dropped by
rescaling $\tilde{Z}(w)$. Clearly $I$ is the ordinary, well-known
linear invariant for the system \cite{hietarinta:secinv}.

\section{Classification of Hamiltonians admitting a cubic
  invariant at arbitrary energy}\label{section:classham}

In this section we present our nontrivial solutions to the
equations \refeq{A2} - \refeq{A0} and \refeq{B2} - \refeq{B0}. For the
third rank Killing tensor cases the solutions can
be fully represented by the analytic function $S(z)$ and the K\"ahler
potential $\mcl{K}$. The solutions are given in four tables
(\ref{table:S0} - \ref{table:S3}), one for each possible degree of the 
polynomial $S(z)$. For the single Killing vector case, the
analytic function $Z(E,z)$ and the potential $V$ give all
information. This solution is given in table \ref{table:kvinv}.

For all cases we will present the potential $V$
(up to linear transformations of the coordinates and a rescaling of the
potential itself) as well as the physical cubic invariant $I$, thereby
making it possible to directly compare our results with Hietarinta's
classification of 1982 \cite{hietarinta:secinv}. A real scaling of
$S(z)$ or $Z(E,z)$ only results in an irrelevant scaling of the
invariant $I$ by the same factor. Accordingly this freedom
will be fixed such that $I$ takes a convenient form. 

For the cases which are superintegrable with
both quadratic and cubic invariant we also indicate, using the
notation of \cite{hietarinta:secinv}, which of the real quadratic
cases (1), (2), (4) or (7) a given system belongs to.

\begin{table}[htbp]
\centering
\begin{tabular}{|l|}\hline
      $S(z) = 4$  \\
      $\mcl{K} = E(x^2+y^2) - f_1(y+\sqrt3\,x) - f_2(y-\sqrt3\,x) -
      f_3(-2y), \quad F_i(\xi) := f_i{}''(\xi) \in \{
      e^{\xi}, \;\mcl{P}(\xi), \;\xi^{-2}, \;\sinh^{-2}\xi,
      \;\sin^{-2}\xi \}$  \\
      $V = F_1(y+\sqrt3\,x) + F_2(y-\sqrt3\,x) + F_3(-2y)$  \\
      $I = p_x{}^3-3p_xp_y{}^2 +
      3[F_1(y+\sqrt3\,x)+F_2(y-\sqrt3\,x)-2F_3(-2y)]p_x +
      3\sqrt3[-F_1(y+\sqrt3\,x)+F_2(y-\sqrt3\,x)]p_y$   \\
      $F_i(\xi)=\xi^{-2}$ is superintegrable, case (2)
      \\ \hline

      $S(z) = 1+i$  \\
      $\mcl{K} = E(x-y)^2-\ts\frac{16}{15}(x^{5/2}\pm y^{5/2})$ \\   
      $V = \sqrt{x} \pm \sqrt{y}$ \\
      $I = p_x{}^3-p_y{}^3+3(\sqrt{x}p_x \mp \sqrt{y}p_y)$ \\
      Superintegrable, case (7) \\ \hline

      $S(z) = 1$  \\
      $\mcl{K} = 2Ey^2-\ts\frac{16}{15}x^{5/2} - 2\delta^{-1}xy -
      \frac{2}{3}\delta y^3$  \\
      $V = \sqrt{x} + \delta y$  \\
      $I = p_x{}^3 + 3\sqrt{x}p_x - \ts\frac32 \delta^{-1} p_y$  \\
      Superintegrable, case (7) \\ \hline

      $S(z) = -1$  \\
      $\mcl{K} = \ts\frac23E(2x^2+y^2)
      -\frac19(8x^4+24x^2y^2-y^4) + 4\delta \ln{\!|y|}$ \\  
      $V = 4x^2+y^2+\delta y^{-2} \quad (\mbox{Drach 1935
      \cite{drach:invcubic}, Holt 1982 \cite{holt:cubinv}})$
      \\  
      $I = p_xp_y{}^2+2(-y^2+\delta y^{-2})p_x + 8xyp_y$ \\
      Superintegrable, case (4) and (7) \\ \hline

      $S(z) = -1$  \\
      $\mcl{K} = \ts\frac23E(2x^2+y^2)
      -\frac23(\frac23x^3+xy^2) + 4\delta \ln{\!|y|}$ \\ 
      $V = x+\delta y^{-2} \quad (\mbox{Drach 1935
        \cite{drach:invcubic}, can be linearly combined with
        the potential above})$ \\
      $I = p_xp_y{}^2+2\delta y^{-2}p_x + yp_y$  \\
      Superintegrable, case (4) and (7) \\ \hline

      $S(z) = -1$  \\
      $\mcl{K} = 2E(2x^2-y^2) + \ts\frac{27}{14}y^{10/3} -
      9(x^2+\delta)y^{4/3}$  \\ 
      $V = \ts\frac34y^{4/3} + (x^2+\delta)y^{-2/3} \quad \mbox{(Drach 
        1935 \cite{drach:invcubic}, Holt
        1982 \cite{holt:cubinv})}$  \\
      $I = 2p_x{}^3 + 3p_x p_y{}^2 +
      3[-3y^{4/3}+2(x^2+\delta)y^{-2/3}]p_x + 18xy^{1/3}p_y$  \\
      \hline

      $S(z) = -1$  \\
      $\mcl{K} = 2E(2x^2-y^2) - 9xy^{-4/3}$  \\
      $V = xy^{-2/3}  \quad (\mbox{Drach 1935 \cite{drach:invcubic},
        can be linearly combined with the 
        potential above})$ \\
      $I = 2p_x{}^3 + 3p_x p_y{}^2 + 6xy^{-2/3}p_x + 9y^{1/3}p_y$  \\
      \hline
\end{tabular}
\caption{Systems for which $S(z)$ is of the zeroth
  degree.}\label{table:S0} 
\end{table}

\begin{table}[htbp]
\centering
\begin{tabular}{|l|}\hline
      $S(z) = -iz$  \\ 
      $\mcl{K} = \ts\frac12E(3x^2+y^2) - \frac{27}{10}x^4 - \frac95
      x^2y^2 - \ts\frac{1}{30}y^4$  \\
      $V = 9x^2+y^2 \quad \mbox{(Fokas and Lagerstr\"om 1980
        \cite{fokas:quadcub}) }$ \\  
      $I = (xp_y-yp_x)p_y{}^2+\ts\frac23 y^3p_x - 6xy^2p_y$  \\
      Superintegrable, case (7) \\ \hline

      $S(z) = -z$  \\
      $\mcl{K} = E(x^2+y^2) - \ts\frac15 A(x^4+4x^2y^2+y^4) +
      4(B\ln{\!|x|}+C\ln{\!|y|})$   \\      
      $V = A(x^2+y^2)+Bx^{-2}+Cy^{-2}$ \quad \mbox{(Drach 1935
        \cite{drach:invcubic})} \\
      $I = (xp_y-yp_x)(p_xp_y+2Axy) - 2Byx^{-2}p_y + 2Cxy^{-2}p_x$  \\
      Superintegrable, case (2) and (7) \\ \hline

      $S(z) = 2iz$  \\  
      $\mcl{K} = Er^2 -
      \ts 6\left\{\sin{\!(2\phi/3)}
      \int{\cos{\!(2\phi/3)}[\cos{\!(2\phi)}]^{-2/3}d\phi} -   
      \cos{\!(2\phi/3)}
      \int{\sin{\!(2\phi/3)}[\cos{\!(2\phi)}]^{-2/3}d\phi}\right\} r^{2/3}$
      \\
      $V = (x^2-y^2)^{-2/3} \quad
      \mbox{(Fokas and Lagerstr\"om 1980 \cite{fokas:quadcub})}$  \\
      $I = (xp_y-yp_x)(p_x{}^2-p_y{}^2) - 4(yp_x+xp_y)(x^2-y^2)^{-2/3}$
      \\  \hline

      $S(z) = 2iz$  \\  
      $\mcl{K} = E(x^2+y^2) - \frac1{10}A(3x^4+2x^2y^2+3y^4)
      -\frac12B\ln{|\,(x+y)\,(x-y)^{-1}|} + C\ln{|x^2-y^2|}$    \\            
      $V = A(x^2+y^2) + [Bxy + C(x^2+y^2)](x^2-y^2)^{-2}$ \\
      $I = (xp_y-yp_x)(p_x{}^2-p_y{}^2) + 2A(xp_y-yp_x)(x^2-y^2)$ \\
      $\quad - [x(x^2+3y^2)(Bp_x+2Cp_y)+y(3x^2+y^2)(2Cp_x+Bp_y)](x^2-y^2)^{-2}$
      \\
      Superintegrable, case (2) \\  \hline
\end{tabular}
\caption{Systems for which $S(z)$ is of the first
  degree.}\label{table:S1} 
\end{table}

\begin{table}[htbp]
\centering
\begin{tabular}{|l|}\hline
      $S(z) = -z^2$  \\ 
      $\mcl{K} = \ts\frac25 E(2x^2+3y^2) - 4[Ar - B\ln{\!(r+x)} -
      C\ln{\!(r-x)}]$  \\     
      $V = [A+B(r+x)^{-1}+C(r-x)^{-1}]r^{-1}$  \\
      $I = (xp_y-yp_x)^2p_x + 2(B+C)[1+(x/y)^2]p_x +
      \left\{-Ay(xp_y-yp_x) + (C-B)\left( \, [2(x/y)^2+3]xp_x+yp_y\,
        \right) \right\}r^{-1}$  \\
      Superintegrable, case (2) and (4) \\  \hline

      $S(z) = -z^2$  \\
      $\mcl{K} = \ts E[1-\frac15\cos{\!(2\phi)}]r^2 - 4(\int\!f(\phi)\,d\phi - C\ln{r})$,\\
      \qquad \mbox{condition:  }
      $[3f''f'-2f'f-Cf'']\sin\phi + [f''f+4(f')^2-2Cf']\cos\phi = 0$\\ 
      $V = f'(\phi)r^{-2}$ \\
      $I = p_\phi{}^2(\cos\phi\,p_r - \sin\phi\,r^{-1}p_\phi) +
      [2f'(\phi)\cos\phi - f(\phi)\sin\phi\,]p_r - [3f'(\phi)\sin\phi +
      f(\phi)\cos\phi\,]r^{-1}p_\phi$  \\  
      Superintegrable, case (2) \\  \hline

      $S(z) = -z^2$  \\  
      $\mcl{K} = \ts\frac25E(2x^2+3y^2) - 4(Ar +
      B\ln{|\,y\,(r-x)^{-1}|} - C\ln{|y|} ) \quad \mbox{(special case of the
        above when $A=0$)}$  \\  
      $V = Ar^{-1}+(Bxr^{-1}+C)y^{-2}$ \quad \mbox{(Drach 1935
        \cite{drach:invcubic})} \\
      $I = (xp_y-yp_x)^2p_x - A(xp_y-yp_x)yr^{-1} +
      [B(2x^2+3y^2)xr^{-1}+2Cr^2]y^{-2}p_x + Byr^{-1}p_y$ \\
      Superintegrable, case (2) \\  \hline
\end{tabular}
\caption{Systems for which $S(z)$ is of the second
  degree.}\label{table:S2} 
\end{table}

\begin{table}[htbp]
\centering
\begin{tabular}{|l|}\hline
      $S(z) = -iz^3$  \\
      $\mcl{K} = Er^2 - 4f(\phi)r^{-1}, \quad \mbox{condition:  }
      f'''f'' - 2f''f' - 3f'f = 0$     \\           
      $V = [f(\phi)+f''(\phi)]r^{-3} \quad \mbox{(compare with Thompson 
        1984 \cite{thompson:polyconst})}$ \\
      $I = p_\phi{}^3 + 3[f'(\phi)p_r + f''(\phi)r^{-1}p_\phi]$  \\
      \hline

      $S(z) = -iz^3$  \\
      $\mcl{K} = Er^2 - A[(\ln{r})^2+\phi^2] - (Be^{\sqrt3 \phi} +
      Ce^{-\sqrt3 \phi})r^{-1} \quad \mbox{(special case of the above
        when $A=0$)}$  \\       
      $V = Ar^{-2} + (Be^{\sqrt3 \phi} + Ce^{-\sqrt3 \phi})r^{-3}$
      \quad \mbox{(Drach 1935 \cite{drach:invcubic}, Hietarinta 1986
        \cite{hietarinta:secinv})} \\ 
      $I = p_\phi{}^3 + 
      \frac{3\sqrt3}4(Be^{\sqrt3\phi}-Ce^{-\sqrt3\phi})p_r +
      \frac34[2A + 3(Be^{\sqrt3\phi}+Ce^{-\sqrt3\phi})r^{-1}]p_\phi$
      \\ \hline
\end{tabular}
\caption{Systems for which $S(z)$ is of the third
  degree.}\label{table:S3} 
\end{table}

\begin{table}[htbp]
\centering
\begin{tabular}{|l|}\hline
      $Z(E,z) = i[Ez-2^{-1/2}(B+iC)\sqrt z]$  \\         
      $V =  (A+B\sqrt{r+x}+C\sqrt{r-x})r^{-1}$ \quad \mbox{(Drach 1935 
        \cite{drach:invcubic})} \\
      $I = (xp_y-yp_x)H + \ts\frac12\left[(B\sqrt{r-x}+C\sqrt{r+x})p_x
       + (-B\sqrt{r+x}+C\sqrt{r-x})p_y\right]$ \\
      Superintegrable, case (4) \\ \hline
\end{tabular}
\caption{A system with a cubic invariant corresponding to a Killing
  vector.}\label{table:kvinv}
\end{table}

\section{Comments}

We have shown that cubic invariants at fixed and arbitrary energy can
be treated in a unified manner by using the Jacobi geometrization
method. Most strongly conserved cubic invariants are nontrivially
cubic in our geometric picture, but we also found a mechanism by which
a new type of cubic invariant can instead correspond to a family of
linear invariants parametrized by the energy. It is then possible to
obtain a cyclic variable by making a standardizing conformal
transformation which is energy dependent. A natural first extension of
this result could be to investigate if the same mechanism can be
explored to find nontrivial quartic invariants which analogously
correspond to quadratic invariants standardized by means of energy
dependent conformal transformations. This is particularly interesting
considering that the standardizing conformal transformations in the
quadratic case are associated with explicit Hamiltonian separability
\cite{rp:inv}. Compare also recent work by Rauch-Wojciechowski and
Tsiganov \cite{rauch:quasisep} who gave some 
examples of non-standard separability. In fact they considered a more
general situation with the separating transformation involving the second
invariant itself in addition to the energy. 

It would also be of interest to apply the method
used in this paper to the usual type of quartic invariants which does
not reduce to quadratic invariants when fixing the energy. In
\cite{larsam:k4} it was shown that the integrability 
condition for fourth rank Killing tensors is of the same nonlinear 
type as in the third rank case. Thus it should be possible to impose
the arbitrary energy condition using the approach adopted in the
present work. 

\section*{Acknowlegments}

We would like to thank Dr.\ A.\ V.\ Tsiganov for helpful remarks.

\end{document}